\documentstyle[12pt,epsf]{article}
\def\beq{\begin{equation}}   \def\eeq{\end{equation}}

\begin{document}
\begin{titlepage}

\begin{flushright}
TPI-MINN-99/37-T\\
UMN-TH-1812/99\\
ITEP-TH-35/99\\
hep-th/9909015
\end{flushright}

\vspace{0.3cm}

\begin{center}
\baselineskip25pt

{\Large\bf
More on the Tensorial Central Charges in ${\cal N} =1$
Supersymmetric
Gauge Theories (BPS Wall Junctions and Strings)}

\end{center}

\vspace{0.3cm}

\begin{center}

{\large A. Gorsky$^{a,b}$ and M. Shifman$^a$}

\vspace{0.2cm}
$^a$ {\em Theoretical Physics Institute, University of Minnesota,
Minneapolis,
MN 55455}

\vspace{0.1cm}

$^b$ {\em Institute of Theoretical and Experimental Physics,
Moscow 117259, Russia$^\dagger$}

\vspace{1cm}

{\large\bf Abstract}

\vspace*{.25cm}

\end{center}

We study  the central extensions of  the ${\cal N} =1$
superalgebras relevant to the soliton solutions
with the axial geometry -- strings, wall junctions, etc.
A general expression valid in any
four-dimensional gauge theory is
obtained. We prove that the only gauge theory
admitting BPS strings at {\em weak coupling} is
supersymmetric electrodynamics with the
Fayet-Iliopoulos term. The problem of ambiguity of
the $(1/2,1/2)$ central charge in the generalized Wess-Zumino
models and gauge theories with matter is addressed and solved.
A possibility of existence of the  BPS strings   at strong
coupling in ${\cal N}=2$ theories is discussed.
A  representation of
different strings within the brane picture is presented.

\vspace{6cm}

$^\dagger$ Permanent address.

\end{titlepage}

\section{Introduction}

In the last several years much has been said about the  domain
walls in various supersymmetric field theories in four dimensions
\cite{1}. The existence of the BPS saturated domain walls is in
one-to-one
correspondence with the central extension of ${\cal N} =1$
superalgebra, with the
central charge $Z_{\alpha\beta}$ lying
in the
representation $\{0,1\}$ or $\{1,0\}$ of the Lorentz
group (for brevity we will
refer  to such charges as the $(1,0)$ charges).
In the non-Abelian gauge theories the $(1,0)$ central charge
emerges as a   quantum  anomaly
in the superalgebra \cite{two} -- \cite{four}.
The possibility of the
existence of the tensorial  central charges  in ${\cal N} =1$
superalgebras was noted in the
brane context in Ref. \cite{deAzcarraga:1989gm}.
The general theory of
the
 central charges in ${\cal N} =1$
superalgebras was revisited recently \cite{FP}.

In this paper we will discuss, in various theories, the
 central extensions of ${\cal N}
=1$ superalgebras with the central charge $Z_{\alpha\dot\beta}$
lying
in the
representation $\{1/2,1/2\}$  of the Lorentz
group (to be referred to as the (1/2,1/2) charges).
Such central charges are related to BPS objects with the axial
geometry,
in particular,
the saturated strings. The fact that
they exist is very well known in the
context of
supersymmetric QED (SQED) with the Fayet-Iliopoulos term,
see Ref. \cite{2,2prim} and
especially Ref. \cite{3}, specifically devoted
to this
issue. In Ref. \cite{3} it is shown, in particular,
that if the spontaneous breaking of U(1) is due to the superpotential
(the so-called $F$ model), then the Abrikosov strings cannot be
saturated. At the same time, if the spontaneous breaking of U(1) is
due
to the Fayet-Iliopoulos term (the so-called $D$ model, with the
vanishing superpotential) then the Abrikosov string is saturated, one
half of supersymmetry is conserved, and the string  tension is given
by the value of the central charge.
\footnote{The statements above refer to ${\cal N}=1$ theories.
In certain ${\cal N}=2$ extensions of QED
one finds BPS saturated strings
without the Fayet-Iliopoulos term. See Secs. 4 and 9.3.}

Another physically interesting
example where  the (1/2,1/2) charges
play a role is the wall junction.
The fact that  generalized Wess-Zumino  (GWZ) models
with a global symmetry of the U(1) or $Z_N$
type may contain BPS wall junctions
was noted in Ref. \cite{AT}.
The interest to the wall junctions  preserving one quarter of the
original supersymmetry was revived recently
after the publications~\cite{wjone,3dwj}, discussing such
junctions
in some GWZ models.

In this work  we calculate the central
extension of the ${\cal N}=1$ superalgebra
of the $Z_{\alpha\dot\beta}$ type for a
generic gauge theory, with or without
matter. As will be seen,
 a spatial integral of a
full  spatial derivative of the
appropriate structure does indeed emerge.
It will be explained how the mass of the saturated solitons with the
axial geometry depends  on the combination of
the $(1,0)$ and $(1/2,1/2)$ central charges. For the solitons that are
pure BPS strings
(i.e. they posses axial geometry, {\em and} their energy density is
completely localized near some axis)  only the $(1/2,1/2)$  charge
can  contribute.
We found that in the Wess-Zumino models, as well as in
the gauge theories with matter, the expression for
this central charge {\em per se} contains certain terms with
coefficients
which are ambiguous. Of critical importance
is the ambiguity in the coefficient of the squark term.
 Using this ambiguity, we
will prove that in {\em weak coupling}
the only ${\cal N}=1$
gauge model admitting the BPS strings is SQED with the
Fayet-Iliopoulos term.  We then present some speculative ideas as to
the possibility of the BPS strings in the non-Abelian models in
{\em strong coupling}. For the objects of the type of the wall
junctions
the ambiguity mentioned above conspires with a related ambiguity
in the $(1,0)$ central charge, so that the resulting energy
of the wall junction configuration is unambiguous.

\section{Generalities}

Let $Q_\alpha\, , \bar Q_{\dot\alpha}$ be supercharges of
the ${\cal N} =1$ four-dimensional field theory under consideration.
The central charge relevant to  strings, $Z_{\alpha\dot\alpha}$,
appears in the anticommutator
\begin{eqnarray}
\{Q_\alpha\, , \bar Q_{\dot\alpha}\}
&\!\! = &\!\!2 P_{\alpha\dot\alpha}
+2 Z_{\alpha\dot\alpha} \nonumber\\[0.2cm]
&\!\!\equiv&\!\!
2\left\{ P_\mu + \int\, {\rm d}^3 x\, \varepsilon_{0\mu\nu\chi}\,
\partial^\nu a^\chi \right\}
\left(\sigma^\mu \right)_{\alpha\dot\alpha}\, ,
\label{bsa}
\end{eqnarray}
where $P_\mu$ is the momentum operator, and $a^\nu$ is an
axial vector  specific to the theory under consideration.
It must be built of
dynamical fields of the theory.
In other words, the $(1/2,1/2)$ central charge is
\beq
Z_\mu = \int\, {\rm d}^3 x\, \varepsilon_{0\mu\nu\chi}
\, \partial^\nu a^\chi\,.
\label{bsaa}
\eeq
The corresponding tensor current
$$
j_{\rho\mu} = \varepsilon_{\rho\mu\nu\chi}
\, \partial^\nu a^\chi
$$
is obviously conserved nondynamically, irrespective of the
concrete form of the axial current $ a^\chi$.

Assume that the string is
aligned along the vector  $n_\mu$ (it is normalized by the condition
$n_\mu n^\mu =-1$), and $L$ is the length of the string ($L$ is
assumed
to tend to infinity).
Then the second term in Eq. (\ref{bsa}) can be always represented as
\beq
Z_\mu = \int\, {\rm d}^3 x\, \varepsilon_{0\mu\nu\chi}
\, \partial^\nu a^\chi  = T L \, n_\mu\, ,
\label{bsaaprim}
\eeq
where $T$ is a parameter of dimension mass squared.
The
direction
of $n_\mu $ can always be chosen in such a way as to make
$T$ in Eq. (\ref{bsaa})  positive. We will always assume $T
>0$.

In the rest frame of the string lying  along the $z$ direction
(i.e. ${\bf n} = \{0,0,1\}$,
or $ n_\mu = \{0,0,0,-1\}$) the superalgebra
(\ref{bsa})
takes the form
\begin{equation}
\{Q_\alpha\, , \bar Q_{\dot\alpha}\} = 2\,  \left[ \begin{array}{cc}
M-T L & 0\\  0 & M+ T L
\end{array} \right]_{\alpha\dot\alpha}\, ,
\label{ququ}
\end{equation}
where $M$ is the total mass of the string.
For the saturated strings
\beq
M = T L\, ,
\eeq
i.e. the mass of the string coincides with the central charge
appearing
in the ${\cal N} =1$ superalgebra (\ref{bsa}). The parameter
$T$ is then identified with the string tension. If the
state of the BPS string is denoted $|\, {\rm str}\rangle$, then
\beq
Q_1|\, {\rm str}\rangle = \bar Q_{\dot 1}|\, {\rm str}\rangle= 0\, .
\eeq
In other words, $Q_1$ and $\bar Q_{\dot 1}$ annihilate the string --
this half of supersymmetry is conserved in the saturated string
background. The action of $Q_2$ and $\bar Q_{\dot 2}$ on $|\, {\rm
str}\rangle$ produces the fermion zero modes.

Any four-dimensional ${\cal N}=1$
theory can be dimensionally reduced
to two dimensions, where it becomes  ${\cal N}=2$ theory.
If the latter has topologically stable instantons,
elevating the theory back to four dimensions gives us strings.
Classical descriptions are totally
equivalent. Distinctions occur at the
level of quantum corrections, which are to be treated differently
in two- and four-dimensional theories. The topological charge of the
two-dimensional theory is related to the central charge of the
centrally
extended algebra (\ref{bsa}). This simple observation
allows one to use a wealth of information regarding various
two-dimensional models
in analysis of saturated strings
in four dimensions at the classical level.

For the solitons of the wall junction type, which preserve
a quarter of the original supersymmetry
(more generally, for the BPS solitons
with the axial geometry), it is  necessary to consider, simultaneously,
the $(1,0)$ charge, which appears in the commutator
\begin{equation}
\{ Q_\alpha Q_\beta\} = -4 i \left(\vec\sigma\right)_{\alpha\beta}\,
\int\, {\rm d}^3 x\,\vec\nabla\, \bar \Sigma\, ,
\label{moycc}
\end{equation}
where $\bar\Sigma$ is a scalar operator built of the dynamical fields
of
the
theory, and
\begin{equation}
 \left(\vec\sigma\right)_{\alpha\beta}=\{-\tau_3\,,\, i\,,\,
\tau_1\}_{\alpha\beta}\,.
\end{equation}

For the BPS strings the $(1,0)$ charge must vanish; however,
for the wall junctions and other axial geometry BPS solitons
both the $(1,0)$ and $(1/2,1/2)$  charges do not vanish (see Sec. 3).
In this case the general structure of the supercharge
anticommutators is as follows
\beq
\frac{1}{2L}\{{\cal Q}\, {\cal Q}\}\to
\begin{array}{c|c|c|c|c}
~~~ & ~\bar Q_{\dot 1}~ & ~\bar Q_{\dot 2}~ & ~ Q_{ 1}~ & ~Q_{ 2}~
\\[0.4cm]\hline
\vspace*{-0.2cm}
Q_{ 1}~ & \frac{M}{L} + \oint a_k {\rm d} x_k & ~~0~~ &
-2i\oint {\rm d}n_k S_k & ~~0~~\\[0.2cm]
Q_{ 2}~ & ~~0~~ & \frac{M}{L} - \oint a_k {\rm d} x_k & ~~0~~ &
~~0~~\\[0.2cm]
\bar Q_{\dot 1}~ & 2i\oint {\rm d}n_kS_k  & ~~0~~ & \frac{M}{L} +
\oint a_k {\rm d} x_k & ~~0~~\\[0.2cm]
\bar Q_{\dot 2} & ~~0~~ &  ~~0~~ &  ~~0~~ &  \frac{M}{L} - \oint a_k
{\rm
d} x_k
\end{array}
\eeq
where the integrals above are taken in the plane perpendicular to
the axis of the soliton
(i.e. in the  $x,y$ plane), along a closed path of
radius $R$
(it is assumed that $R\to\infty$),
 $dn_k$ is the element of the length of the curve, see Fig. 1 ($ d\vec
n$ is
perpendicular
to $ d\vec x$), and, finally,
\beq
\{ S_1\,,\, S_2\} = \{ {\rm Re}\Sigma \,,\, {\rm
Im}\Sigma\}\, ,
\eeq
so that
\beq
\oint a_k {\rm d} x_k = \int {\rm d}^2 x (\partial_x a_y - \partial_y
a_x
)
=\int {\rm d}^2 x \left[-i\partial_\zeta (a_x +ia_y) +
i\partial_{\bar\zeta}
(a_x -ia_y)\right]\, ,
\eeq
\beq
\oint {\rm d}n_k S_k = \int {\rm d}^2 x\left[\partial_\zeta \Sigma
+\partial_{\bar\zeta}\bar\Sigma \right]\, ,
\eeq
and the complex coordinates $\zeta, \bar\zeta$ are introduced below
in Eq. (\ref{zetabar}).
The BPS bound on the
soliton mass is
obtained from the requirement of vanishing of the determinant of
the
above matrix, which implies
\beq
 \frac{M}{L} =  -\oint a_k {\rm d} x_k + 2\oint {\rm d}n_k S_k
\, .
\label{ccbpss}
\eeq
For saturated objects
the master equation (\ref{ccbpss})
expresses the tensions in terms of two contour integrals
over the large circle.

\begin{figure}
\epsfysize=6cm
\centerline{\epsfbox{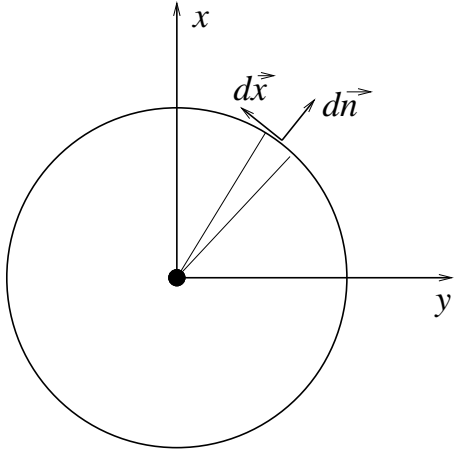}}
 \caption{The integration contour in the $x,\, y$ plane. The soliton
axis
(the closed circle) lies perpendicular to this plane.}
\end{figure}

\section{Generalized Wess--Zumino Models}

In this section, as a warm up exercise,
 we will discuss the GWZ models which give rise to the
BPS solitons with the axial geometry, and derive the $(1/2,1/2)$
central charge in these models. The full expression for the
$(1,0)$ central charge was found previously \cite{four}.
The Lagrangian has the form
\begin{equation}
{\cal L} =
\frac{1}{4} \sum_i \int \! {\rm d}^2\theta {\rm d}^2\bar\theta\,
\bar \Phi_i  \Phi_i +\left\{\frac{1}{2}\int \! {\rm d}^2\theta
{\cal W} (\Phi_i ) + \mbox{H.c.}
\right\}
\, ,
\label{GWZlagr}
\end{equation}
where $\Phi_i$ is the set of the chiral fields, and the superpotential
${\cal W}$ is an analytic function of the fields $\Phi_i$.
The original (renormalizable) Wess--Zumino model
implies that ${\cal W}$ is a cubic polynomial in $\Phi_i$.
We shall not limit ourselves  to this assumption, keeping in  mind
that
GWZ models with more contrived
superpotentials can appear as low-energy limits of some
renormalizable microscopic field theories. The case
of more general K\"ahler potential will
be considered later.

The equations of the BPS saturation
for the solitons with the axial geometry
 in this model were first derived~\footnote{
See Sec. III.D of Ref.~\cite{four} entitled, rather awkwardly,
``BPS-saturated strings." In fact, the
authors meant BPS solitons with the
axial
geometry.}
 in Ref.~\cite{four}; they have the form
\begin{equation}
\frac{\partial \phi_i }{\partial\zeta} = \frac{1}{2}\frac{\partial
\bar{\cal
W}}{\partial
\bar\phi_i}\, ,
\label{Aspenone}
\end{equation}
where
\begin{equation}
\zeta = x+i y\,,\quad \frac{\partial }{\partial\zeta}=\frac{1}{2}
\left( \frac{\partial }{\partial x}- i\frac{\partial }{\partial y} \right)\,
.
\label{zetabar}
\end{equation}
The soliton axis is assumed to lie along the $z$ axis,
while the soliton profile depends on $x,y$. Note that it is {\em not}
assumed that the solution of Eq. (\ref{Aspenone})
is analytic in $\zeta$ (in fact, one can prove that
  it must depend on both $\zeta$ and $\bar\zeta$ in the general
case).
A constant phase, which could have appeared on the right-hand side
of
Eq. (\ref{Aspenone}), is absorbed in $\zeta$.

Given the solution of Eq. (\ref{Aspenone}), one gets two constraints
determining  the parameter of the residual (conserved)
supersymmetry,
\begin{equation}
 \left( 1 + \tau_3\right) \varepsilon = 0\, ,\quad
\frac{-i}{2} \left( 1 - \tau_3\right) \varepsilon = \bar\varepsilon
\, ,
\label{prs}
\end{equation}
where the spinorial indices of $\varepsilon,\bar\varepsilon$
are suppressed (both are assumed to be the upper indices),
and we follow the notations and conventions collected in
\cite{four}. The first constraint implies that $\varepsilon$ has only
the
lower
component, which reduces the number of supersymmetries from
four to
two;
the second constraint further reduces the number of the
residual supersymmetries to one.

In order to calculate the $(1,0)$ and $(1/2,1/2)$
central charges one needs the expression for the supercharges. In
fact,
since we  focus on full derivatives, we need to know the
supercurrent
$J^\mu_\alpha =(1/2)\left(\bar\sigma\right)^{\dot\beta\beta}
J_{\alpha\beta\dot\beta}$, rather than the
supercharges {\em per se}. The corresponding expression is well
known
(see e.g. Ref. \cite{four}),
\begin{eqnarray}
J_{\alpha\beta\dot\beta} &\!\!=&\!\!   2\sqrt{2}\sum
\left[
\left({\partial
}_{\alpha\dot\beta}\bar\phi
\right)\psi_\beta -i\epsilon_{\beta\alpha} F\bar\psi_{\dot\beta}
\right]
\nonumber\\[0.2cm]
&\!\!- &\!\! \frac{\sqrt{2}}{3}\sum \left[
\partial_{\alpha\dot\beta}(\psi_\beta \bar\phi)
+\partial_{\beta\dot\beta}(\psi_\alpha \bar\phi)
-3\epsilon_{\beta\alpha}\partial^\gamma_{\dot\beta}(\psi_\gamma
\bar\phi)
\right]
\, .
\label{scwz}
\end{eqnarray}
The supercharge $Q_\alpha$ is defined as
\beq
Q_\alpha =\int d^3 x J_\alpha^0\, , \qquad
J^\mu_\alpha = \frac{1}{2} \left(
\bar\sigma^\mu\right)^{\beta\dot\beta}
J_{\alpha\beta\dot\beta}\, .
\label{oprsz}
\eeq
The term in the second line in Eq. (\ref{scwz}) is conserved by itself.
Moreover,
in the supercharge it is represented as an integral over the full
derivative.
Below we will discuss the impact of deleting this term.
We will keep it, however, for the time being,
since we  want to use  the supercurrent which enters in one
supermultiplet
with the geometric $R$ current \cite{scsm} (sometimes called the
$R_0$ current). The $R_0$ current is conserved in conformal
theories.

It is not difficult to find the full derivative terms  in $\{Q_\alpha
\bar Q_{\dot\beta}\}$ by computing the canonic commutators of the
fields at the
tree level [the $(1/2,1/2)$ central charge appears already at the tree
level].
The task is facilitated if one observes that in order to get the
$(1/2,1/2)$ central
charge it is sufficient to keep only the terms of the mixed symmetry
in
$\{\bar Q_{\dot\alpha}\,
J_{\alpha\beta\dot\beta}\}$, namely, symmetric in $\alpha ,\beta$
and
antisymmetric in $\dot\alpha ,\dot\beta$ or {\em vice versa}.

The result of this calculation reduces to Eq. (\ref{bsaa})
with
\begin{equation}
a^\mu = \frac{1}{4} \, a^\mu_{(\psi )}- \frac{1}{6} a^\mu_{(\phi )}\,  ,
\end{equation}
where $a^\mu_{(\psi )}$ and $ a^\mu_{(\phi )}$ are the fermion and
boson axial
currents, respectively,
\begin{equation}
 a^\mu_{(\psi )} = -\sum \psi \sigma^\mu \bar \psi\, ,
\qquad
 a^\mu_{(\phi )}= - i\sum \phi
\stackrel{\leftrightarrow}{\partial}^{\,\mu}  \bar\phi \,  .
\end{equation}

The expression for the $(1,0)$ central charge in the GWZ model
found previously \cite{four} at the tree level takes the form of Eq.
(\ref{moycc})
with
\beq
\bar\Sigma = \bar{\cal W} - \frac{1}{3}\sum\, \bar\Phi\,
\frac{\partial
{\cal W}}{\partial
\bar\Phi}\,.
\eeq

One can check that  {\em only} the {\em combined} contribution
 of the central charges above
correctly reproduces the mass of the BPS solitons
with the axial geometry, e.g. the wall junctions.
Indeed, Eq. (\ref{ccbpss}) implies that in the model at hand
\footnote{The term $-(1/3)\partial_k\partial_k (\bar\phi\phi)$ is
irrelevant both
for strings and wall junctions, since it vanishes
in the both cases. It contributes, however, in the energy  of the
axial geometry solitons of the type discussed in \cite{four}.  This
term
occurs in
passing from the canonic energy-momentum tensor
$$
\theta_{\mu\nu}^{\rm canonic} = \partial_\mu\bar\phi
\partial_\nu\phi +
\partial_\nu\bar\phi  \partial_\mu\phi +{\rm fermions}
- g_{\mu\nu}{\cal L}
$$
to the one which is traceless in the conformal limit
$$
\theta_{\mu\nu}^{\rm traceless} = \theta_{\mu\nu}^{\rm canonic} +
\frac{1}{3}\,\left( g_{\mu\nu} \partial^\alpha\partial_\alpha
-\partial_\mu\partial_\nu
\right)\bar\phi\phi\,.
$$
}
\begin{eqnarray}
\frac{M}{L}&\!\!=&\!\!  \int {\rm d}^2 x\left[\partial_k\bar\phi
\partial_k\phi
 +\left|\frac{\partial W}{\partial\phi}\right|^2
-\frac{1}{3}\partial_k\partial_k (\bar\phi\phi)
\right]\nonumber\\[0.2cm]
&\!\!=&\!\! -2\left( 1 -\frac{2}{3}\right)\int {\rm d}^2 x\left[
\partial_\zeta\phi \partial_{\bar\zeta}\bar\phi
-\partial_\zeta\bar\phi
\partial_{\bar\zeta}\phi
\right]
\nonumber\\[0.2cm]
&\!\!+&\!\! 2\int {\rm d}^2 x\left[\partial_\zeta\left({\cal W}
-\frac{1}{3}\phi
\frac{\partial {\cal W}}{\partial\phi}\right)
+\partial_{\bar\zeta}\left(\bar{\cal W} -
\frac{1}{3}\bar\phi
\frac{\partial \bar{\cal W}}{\partial\bar\phi}\right)
\right] \, .
\label{smccp}
\end{eqnarray}

On the other hand, for the BPS-saturated solution one can write
\begin{eqnarray}
0&\!\!=&\!\!  \int {\rm d}^2 x\left[2\partial_\zeta\phi-
\frac{\partial
\bar{\cal
W}}{\partial\bar\phi}\right]
\left[2\partial_{\bar\zeta}\bar\phi- \frac{\partial {\cal
W}}{\partial\phi}\right]
\nonumber\\[0.2cm]
&\!\!=&\!\! \int {\rm d}^2 x\left[\partial_k\bar\phi \partial_k\phi
 +\left|\frac{\partial {\cal W}}{\partial\phi}\right|^2
\right] \nonumber\\[0.2cm]
&\!\!+&\!\! 2\int {\rm d}^2 x\left[
\partial_\zeta\phi \partial_{\bar\zeta}\bar\phi
-\partial_\zeta\bar\phi
\partial_{\bar\zeta}\phi
\right]
- 2\int {\rm d}^2 x\left[\partial_\zeta{\cal W}
+\partial_{\bar\zeta}\bar{\cal W}
\right] \, ,
\label{smbpssp}
\end{eqnarray}
or
\beq
\frac{M}{L} =  -2\int {\rm d}^2 x\left[
\partial_\zeta\phi \partial_{\bar\zeta}\bar\phi
-\partial_\zeta\bar\phi
\partial_{\bar\zeta}\phi
\right]
+ 2\int {\rm d}^2 x\left[\partial_\zeta{\cal W}
+\partial_{\bar\zeta}\bar{\cal W}
\right] \, .
\label{smbpss}
\eeq

At first sight it might seem
that Eqs. (\ref{smccp}) and (\ref{smbpss}) contradict each other,
since the axial current contribution to the soliton mass in these two
expressions
(corresponding to the $(1/2,1/2)$ central charge)
has different coefficients (cf. $-2 +(4/3)$ in the first case and $-2$
 in the
second).
Upon inspection one sees that Eq. (\ref{smccp}) has a different
expression for the
$(1,0)$ central charge too. The difference is
$$
-\frac{2}{3}\int {\rm d}^2 x\left[\partial_\zeta\left(\phi
\frac{\partial {\cal W}}{\partial\phi}\right)
+\partial_{\bar\zeta}\left(
\bar\phi
\frac{\partial \bar{\cal W}}{\partial\bar\phi}\right)
\right] \, .
$$
For the BPS saturated solitons satisfying Eq. (\ref{Aspenone})
it is easy to show that
\begin{eqnarray}
& &-\frac{2}{3}\int {\rm d}^2 x\left[\partial_\zeta\left(\phi
\frac{\partial {\cal W}}{\partial\phi}\right)
+\partial_{\bar\zeta}\left(
\bar\phi
\frac{\partial \bar{\cal W}}{\partial\bar\phi}\right)
\right]\nonumber\\[0.2cm]
&\!\!=&\!\!
-\frac{4}{3}\int {\rm d}^2 x\left[
\partial_\zeta\phi \partial_{\bar\zeta}\bar\phi
-\partial_\zeta\bar\phi
\partial_{\bar\zeta}\phi
\right] -\frac{1}{3} \int {\rm d}^2 x \partial^\alpha\partial_\alpha
\bar\phi\phi\,.
\label{vsprav}
\end{eqnarray}
This relation immediately implies the coincidence of the
soliton masses ensuing from  Eqs. (\ref{smccp}) and (\ref{smbpss}),
respectively.

In fact, the superficial difference between them
is due to the ambiguity in the choice of the
supercurrent (the terms with the full derivatives in Eq. (\ref{scwz}))
and the corresponding ambiguity in the energy-momentum tensor.
Equation (\ref{smccp}) is derived on the basis
of the supercurrent and  the energy-momentum tensor
with the properties
$\varepsilon^{\alpha\beta}J_{\alpha\beta\dot\beta}=0,\,\,\,
\theta^\mu_\mu =0 $
in the conformal limit. Passing to the minimal supercurrent
and the canonic energy-momentum tensor
one drops all terms containing the factor $1/3$ in Eq. (\ref{smccp})
and recovers Eq. (\ref{smbpss}). The mass of the soliton stays intact
due to a reshuffling of contributions due to $(1/2,1/2)$ and $(1,0)$
charges.

To illustrate the point let us consider, for instance, a $Z_N$
model suggested in Ref. \cite{GabDv},
 with the
superpotential
\beq
{\cal W} =N \left\{  \Phi  -
\frac{N}{N+1}\left(\frac{\Phi}{N}\right)^{N+1}\right\} \, ,
\eeq
where  $\Phi$ is a chiral superfield. The model
obviously possesses a $Z_N$ symmetry, the vacuum manifold
corresponds
to $N$ points,
\beq
\phi_k = N \exp\left(\frac{2\pi i k}{N}\right)\,,\qquad k= 0,1,2, ...,N-
1\, ,
\eeq
while the vacuum value of the superpotential is
\beq
{\cal W} (\phi_k) =  N^2 \exp\left(\frac{2\pi i k}{N}\right)\,,\qquad
N\to\infty\, .
\eeq
The solution of the BPS saturation equation for an isolated wall
exists, it was discussed in \cite{GabDv}. (Here and below $N$ will be
assumed large,
and only  leading terms in $N$ will be kept.)
The tension of the minimal wall connecting the neighboring vacua is
\beq
T= 2|\Delta {\cal W}| = 4\pi N\, .
\eeq
Consider the BPS wall junctions of the type depicted in Fig. 2.
\begin{figure}
\epsfysize=6cm
\centerline{\epsfbox{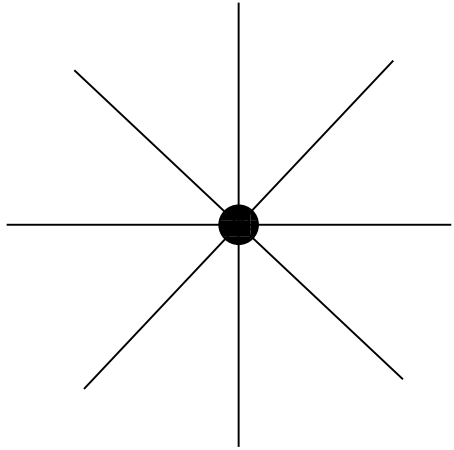}}
 \caption{The domain wall junction in the theory with $Z_N$
symmetry. The
``hub" is denoted by the closed circle.}
\end{figure}
Assuming that there is a solution of Eq. (\ref{Aspenone}),
to the leading order in $N$ one can write (at $|\zeta |\to \infty$)
\beq
\phi = N e^{i\alpha (\gamma )}\, , \quad \alpha (0) = 0\,,\quad
\alpha
(2\pi ) = 2\pi\,,
\eeq
which entails, in turn,
\beq
\oint_{|x|=R\to\infty}\,  a_k {\rm d}x_k =\frac{N^2}{3}\left[ \alpha
(2\pi ) -
\alpha (0 )\right] = \frac{2\pi}{3}N^2\, .
\eeq
We also observe that
\begin{eqnarray}
2\oint {\rm d}n_k w_k &\!\!=&\!\!  2 N^2 R\int {\rm d}\gamma
\cos(\alpha -\gamma )
=4\pi N^2 R\, ,\nonumber\\[0.2cm]
 \{w_1, \,w_2\} &\!\!=&\!\!  \{{\rm Re }{\cal W}\, , \, {\rm Im }{\cal
W}\}\,.
\label{okot}
\end{eqnarray}
which is exactly the mass of $N$ isolated walls inside the contour.
Furthermore,
\beq
2\oint {\rm d}n_k S_k
=4\pi N^2 R -\frac{4\pi}{3}N^2\, .
\label{okotp}
\eeq
The total mass of the junction configuration comes out the same from
both
expressions, Eqs. (\ref{smccp}) and (\ref{smbpss}),
\beq
\frac{M}{L} = 4\pi N^2 R - 2\pi N^2\, ,
\eeq
(see also \cite{GabShi}).
The first term can be interpreted as the mass of the ``spokes"
joined at the origin,
while the second as that of the
``hub".

Let us remark that the stringy (``hub")  contribution to the
total mass equals to twice the area of the contour
on the $\phi$ plane covered by the solution. Since
we consider the junction with $N$  ``minimal" domain walls
connecting the neighboring vacua, the contour is closed.
The closeness is nothing but the equilibrium
condition at the junction line.

Summarizing, we observe an ambiguity in the
$(1/2,1/2)$ central charge.
This ambiguity is due to the
fact that both, the supercurrent and the energy-momentum tensor,
are not uniquely determined. Both admit certain full
derivative terms which are conserved by themselves
and, therefore, do not affect the supercharges
and the energy-momentum four-vector.
They do affect the expressions for the central charges, however.
For the soliton solutions of the wall junction
type the ambiguity in the
$(1/2,1/2)$ central charge combines with another ambiguity,
in the $(1,0)$ central charge, to produce an
unambiguous expression for the soliton mass. As we will see shortly,
the same ambiguity (and a similar conspiracy) takes place in
the gauge theories
with matter.

Practically, it is more convenient to
work with the minimal supercurrents
(and the canonic energy-momentum tensor).
Then, one omits the second line in Eq.
(\ref{scwz}). The expression for $a^\mu$
in the $(1/2,1/2)$ central charge then  becomes
\beq
a^\mu = \frac{1}{4} \, a^\mu_{(\psi )}- \frac{1}{2} a^\mu_{(\phi )}\, ,
\qquad
 a^\mu_{(\phi )}= - i\sum \phi
\stackrel{\leftrightarrow}{\partial}^{\,\mu}  \bar\phi \, ,
\eeq
while $\bar\Sigma$ in the  $(1,0)$ central charge becomes
$$\bar\Sigma = \bar{\cal W}\,.$$

\section{SQED with the Fayet-Iliopoulos term}

The simplest theory (and the only one in the class
${\cal N}=1$, see
below)  where saturated  strings exist  in the weak coupling
regime
is supersymmetric electrodynamics (SUSY QED, or SQED), with the
Fayet-Iliopoulos (FI) term.
In the superfield notation the Lagrangian of the
model has the form
\begin{equation}
{\cal L } = \left\{ \frac{1}{8\, e^2}\int\!{\rm d}^2\theta \, W^2 + {\rm
H.c.}\right\} +
\frac{1}{4}\int \!{\rm d}^4\theta \left(\bar{S}e^V S + \bar{T}e^{-V} T
\right)
 - \frac{\xi }{4} \int\! {\rm d}^2\theta {\rm d}^2\bar \theta
\,V(\! x,\theta , \bar\theta ) \, ,
\label{sqed}
\end{equation}
where $e$ is the electric charge,
 $S$ and $T$ are two chiral superfields with the electric charges
$+1$ and $-1$, respectively, $\xi$ is the coefficient
of the Fayet-Iliopoulos term.   The
model with one chiral superfield is internally anomalous.
Topologically
stable solutions in this model and its modifications were considered
more than once in the past
\cite{2,2prim,3}. We combine various elements scattered in the
literature,
with a special emphasis on the algebraic aspect.  Supersymmetry of
this
model is minimal, ${\cal N} =1$.

If $\xi\neq 0$, the vacuum state corresponds to the spontaneous
breaking of U(1). The spectrum of the model is that of a massive
vector
supermultiplet (one massive vector field, one real scalar and one
Dirac
fermion, all of one and the same mass),
plus a massless modulus (one chiral superfield)
 parametrized by the product $ST$,
\beq
\Phi = 2ST\, .
\eeq
The vacuum valley is represented by the one-dimensional complex
manifold with the K\"ahler function
\beq
K(\Phi\, , \bar\Phi ) = \sqrt{\xi^2 +\Phi \bar\Phi}\, .
\eeq
In a generic point
nonsingular Abrikosov strings do {\em not} exist \cite{2prim}.
There is one special point, however, $\Phi = 0$, where
the theory supports the saturated string.

In components the Lagrangian of SQED (\ref{sqed}) has the form
(in the Wess-Zumino gauge)
\begin{eqnarray}
{\cal L } &\!\!=&\!\!-\frac{1}{4e^2}F_{\mu\nu}F^{\mu\nu}
+\left({\cal D}_\mu\phi\right)^\dagger {\cal D}^\mu\phi
+\left({\cal D}_\mu\chi\right)^\dagger {\cal D}^\mu\chi
-\frac{e^2}{2}\left( \phi^\dagger \phi -
\chi^\dagger\chi -\xi\right)^2\, ,
\nonumber\\[0.2cm]
&\!\!+&\!\!\mbox{fermions}
\label{lqedc}
\end{eqnarray}
where $\phi$ and $\chi$ are the lowest components
of the superfields $S$ and $T$, respectively, with the
electric charges $\pm 1$, e.g.
\beq
{\cal D}_\mu\phi =\partial_\mu \phi - i A_\mu\phi\, , \qquad
[{\cal D}_\mu\, , {\cal D}_\nu ]\phi = -i F_{\mu\nu }\phi\, .
\eeq
Without loss of generality we can assume that $\xi >0 $.

For the static field configurations, assuming in addition that all
fields
 depend only on $x$ and $y$ and $A_0=A_3 = 0$, one gets the energy
functional in the form
\begin{eqnarray}
{\cal E} &\!\!=&\!\!\int{\rm d}x{\rm d}y
\left\{ \frac{1}{2e^2}F_{12}^2
+\sum_{i=1,2}\left({\cal D}_i\phi\right)^\dagger{\cal D}_i\phi +
\frac{e^2}{2}\left(\phi^\dagger\phi
-\xi
\right)^2\right\}\nonumber\\[0.2cm]
&\!\!\equiv &\!\! \int{\rm d}x{\rm d}y
\left\{  \left| \frac{1}{\sqrt{2}e} F_{12} +\frac{e}{\sqrt{2}}
\left(\phi^\dagger\phi -\xi\right)
\right|^2\right.
\nonumber\\[0.2cm]
&\!\!+&\!\!
\left. \left[ \left(
{\cal D}_1+ i{\cal D}_2\right)\phi\right]^\dagger\,
\left(
{\cal D}_1+ i{\cal D}_2\right)\phi \right\}+ {\cal Q}\, ,
\label{sqedef}
\end{eqnarray}
where ${\cal Q}$ is the surface (topological) term,
\beq
{\cal Q} =  \int{\rm d}x{\rm d}y
\left\{ \xi F_{12} - \frac{i}{2} \partial_i (\phi^\dagger
\stackrel{\leftrightarrow}{\cal D}_j\phi )
\varepsilon^{ij}\right\}\, , \qquad i,j = 1,2\, .
\label{tt}
\eeq
We will discuss the value of the surface term later.

The saturation  equations are
\begin{eqnarray}
F_{12} &\!\!=&\!\!  -e^2
\left(\phi^\dagger\phi -\xi\right)\, ,\nonumber\\[0.2cm]
\left(
{\cal D}_1+ i{\cal D}_2\right)\phi &\!\!=&\!\!   0\, .
\label{sateq}
\end{eqnarray}
The {\em Ansatz} which goes through these equations is
\begin{eqnarray}
\phi &\!\!=&\!\!  \sqrt{\xi}\eta e^{i\alpha}
\, ,\nonumber\\[0.2cm]
A_i &\!\!=&\!\!   a\, \frac{\partial\alpha}{\partial x^i}\,, \quad
i=1,2\, ,
\label{agt}
\end{eqnarray}
where
\beq
\alpha = \mbox{Arg}\, \zeta\, , \qquad \zeta = x+i y\,,
\eeq
and $\eta,\,\, a$ are some functions depending on $r$.
This must be supplemented  by the standard boundary conditions,
namely
\beq
\eta (r)\, ,\,\,  a(r) \longrightarrow
\left\{ \begin{array}{c}
0\quad\mbox{at}\quad r\to 0\\  1\quad\mbox{at}\quad r\to \infty
\end{array} \right.\, .
\eeq
For the given {\em Ansatz} the saturation equations (\ref{sateq})
degenerate into a system of first-order equations
\begin{eqnarray}
a ' &\!\!=&\!\!  e^2\xi r (\eta^2 -1)
\, ,\nonumber\\[0.2cm]
\eta ' &\!\!=&\!\!   -\frac{\eta(1-a)}{r} \, ,
\label{sateqa}
\end{eqnarray}
where the prime denotes differentiation over $r$.
Its solution  is well known.

It is instructive to compare the topological term in Eq. (\ref{tt})
with the central charge of the superalgebra.  To derive the central
charge one needs the
expression for the  supercurrent in SQED, which takes the form (in
the
spinorial notation)
\begin{eqnarray}
J_{\alpha\beta\dot\beta} &\!\!=&\!\!  \frac{2}{e^2}
\left(iF_{\beta\alpha} \bar\lambda_{\dot\beta}
+ \epsilon_{\beta\alpha} D\bar\lambda_{\dot\beta}
\right) + 2\sqrt{2}\sum\left({\cal D}_{\alpha\dot\beta}\phi^\dagger
\right)\psi_\beta
\nonumber\\[0.2cm]
&\!\!- &\!\! \frac{\sqrt{2}}{3}\sum \left[
\partial_{\alpha\dot\beta}(\psi_\beta \phi^\dagger)
+\partial_{\beta\dot\beta}(\psi_\alpha \phi^\dagger)
-3\epsilon_{\beta\alpha}\partial^\gamma_{\dot\beta}(\psi_\gamma
\phi^\dagger)
\right]
\, .
\label{scsqed}
\end{eqnarray}
Above it is assumed that there is no superpotential.
The second line may or may not be added, at will.
(The second line in Eq. (\ref{scsqed}) is conserved by itself;
in the supercharge it presents a full spatial derivative, hence, its
contribution
vanishes.)
The sum runs over various matter supermultiplets, in particular, $S$
and $T$ in
the case at hand.

To  find the central charge one must compute the anticommutator
$\{ Q_\alpha \, , \bar J_{\dot\beta\dot\gamma\delta}\}$. Moreover,
we decompose the anticommutator above with respect to irreducible
representations of the Lorentz group, by  singling out the
symmetric and
antisymmetric combinations of the dotted and undotted indices.
The one which is symmetric with respect to both pairs,
$(\alpha , \delta )$ and $(\dot\beta\dot\gamma )$, is the Lorentz
spin 2 (the energy-momentum tensor), which contributes to
$P_{\alpha\dot\alpha}$, rather than to the central charge.
The combination which is antisymmetric with respect to both pairs,
$(\alpha , \delta )$ and $(\dot\beta\dot\gamma )$ is Lorentz singlet,
it represents the trace terms in the energy-momentum tensor.
To single out the central charge we must isolate the terms of the
mixed
symmetry, i.e. symmetric with respect to $(\alpha , \delta )$ and
antisymmetric with respect to $(\dot\beta\dot\gamma )$,
and {\em vice versa}.

Keeping in mind this remark, and using the canonic commutation
relations and equations of motion for the $D$ field we get
an expression similar to that in the Wess-Zumino model,
plus an extra contribution due to the $D$ term,
\beq
\{ Q_\alpha \bar Q_{\dot\alpha}\} = i \xi \int d^3 x
\left[
 F_{\beta\alpha}  \varepsilon_{\beta\dot \alpha}
- \bar F_{\dot\gamma\dot\alpha}
\varepsilon_{\dot\gamma\alpha}
\right]\, .
\eeq
This implies
\beq
Z_\mu = \int {\rm d}^3 x\, \varepsilon_{0\mu\nu\rho}
\left(\xi\,  \partial^\nu A^\rho  -\sum \frac{i}{2}\partial^\nu
(\bar\phi
\stackrel{\leftrightarrow}{\cal D}^{\,\rho}  \phi )
+ \frac{1}{4} \partial^\nu R^\rho
+\frac{1}{4}\partial^\nu a_{(\psi)}^\rho \right)\, ,
\label{tfit}
\eeq
where $R^\rho$ is the photino current, while $a_{(\psi)}^\rho$
is that of the electrons,
\beq
R^\rho = -\frac{1}{e^2}\lambda\sigma^\rho\bar\lambda\,,\qquad
 a^\mu_{(\psi )} = -\sum \psi \sigma^\mu \bar \psi\, .
\eeq

Note that the coefficient of the $\bar\phi
\stackrel{\leftrightarrow}{\cal D}^{\,\rho}  \phi$
(i.e. the selectron axial current) term is ambiguous --
it depends on whether the second line in Eq. (\ref{scsqed}) is
included
in the definition of the supercurrent. The result quoted above refers
to the minimal supercurrent, with the second line in Eq.
(\ref{scsqed})
discarded. Since the $(1,0)$
central charge is irrelevant for the string solution,
this ambiguity alone shows that the $\bar\phi
\stackrel{\leftrightarrow}{\cal D}^{\,\rho}  \phi$
term cannot contribute to the central charge under consideration.
It is certainly the case, since ${\cal D}^{\,\rho}  \phi$
falls off sufficiently fast at $r\to\infty$
(where $r$ is the distance to the string axis) for the string solution.
At the same time, the photon four-potential $A^\rho$
falls off slowly, as $1/r$.
Thus, the $(1/2,1/2)$ central charge is saturated
by the $\xi$ term exclusively. The latter is unambiguously
fixed in Eq. (\ref{tfit}), i.e. it does not depend on the full derivative
terms
in the supercurrent.  The $(1/2,1/2)$ central charge is
obviously proportional to
$\xi$ and to the magnetic flux of the string,
\beq
\frac{M}{L}= \xi{\cal F}\,,
\eeq
where
\beq
{\cal F} =\int {\rm d}x {\rm d}y F_{12} =\oint A_k {\rm d } x_k\,.
\eeq

Note that the very same saturation equations (\ref{sateq})
are obtained in ${\cal N}=2$ SQED
with the vanishing Fayet-Iliopoulos term
and linear superpotential, see Sec. 9.3.

\section{The  K\"ahler Sigma Models}

In this section we present some arguments concerning strings
in the four-dimensional  $\sigma$ models on the  K\"ahler manifolds.
The two-dimensional reductions of these models
are well studied, in the Euclidean formulation they admit instantons,
which are the solutions of the first order self-duality equations.
In the supersymmetric version the self-duality equations
in two dimensions
are reinterpreted as the BPS equations in higher-dimensional
theories (e.g. \cite{dopone}). It is obvious that
the instantons of the two-dimensional models are the BPS strings
in four dimensions. Thus, the four-dimensional
$\sigma$ models on the  K\"ahler
manifolds do have the BPS strings at
the quasiclassical level, at weak coupling.
Keeping in mind the assertion we are going to prove
later (Sec. 8) we discuss where the  K\"ahler sigma models
stand compared to other models.

Let us start with the $CP_1$ model. In a sense, this model can be
obtained as a limiting case of
SQED  with a somewhat  different matter content
compared to that of Sec. 4 (see, for instance,
\cite{wittenn=2}). Indeed, assume that the matter superfields
$S$ and $T$ have
both charges $+1$, rather than $\pm 1$. As a quantum theory, it is
anomalous,
but for the time being we limit ourselves to the classical
consideration. The limit to be taken is $e^2\to\infty$.
Let us have a closer look at Eq. (\ref{lqedc}), with the sign of the
charge
of the $\chi$ field reversed [correspondingly,
the $D$ term takes the form $D =e^2( \phi^\dagger \phi +
\chi^\dagger\chi -\xi)$].
In this limit the photon mass tends to infinity, the photon becomes
nondynamical and can be eliminated. It drags with itself
two real scalar degrees of freedom. The remaining two scalar degrees
of
freedom are massless.  Their interaction reduces to
the sigma model on a sphere.  This is most easily seen from Eq.
(\ref{lqedc}).
In the limit $e^2\to\infty$ the  $D$ term must vanish, which implies
that $ \phi^\dagger \phi +
\chi^\dagger\chi =\xi$.  In fact, the gauge freedom allows one to
identically  eliminate
one out of four degrees of freedom residing in $\phi,\,\, \chi$.
The remaining three are subject to the constraint, telling
us that the radius of the sphere is $\xi$.

Thus, the SQED with the Fayet-Iliopoulos term, in the limit
$e^2\to\infty$,  gives rise to the model with the
 action
\beq
S = \frac{1}{2g^2} \int {\rm d}^4 x  {\rm d}^2\theta
 {\rm d}^2\bar\theta\, \ln \left (1+\bar\Phi\Phi \right)
\eeq
where $\Phi$ is a chiral superfield,
\begin{equation}
{\Phi ({x}_L,\theta )} = \phi ({x}_L) + \sqrt{2}\theta^\alpha
\psi_\alpha ({
x}_L) +  \theta^2 F({x}_L)\, .
\label{chsup}
\end{equation}
The coupling constant $2/g^2$ has the dimension of mass squared
and is equal to $\xi$.
The string tension will be proportional to $2/g^2=\xi$.
The metric of the sphere in the target space $G$ in this case is
\beq
G = \frac{2}{g^2} \frac{1}{(1+\bar\Phi\Phi )^2 }\, .
\eeq
The energy functional for the stringy solution takes the
form which looks exactly as
the action in the Euclidean two-dimensional sigma
model whose world volume is transverse
to the string.  It is easy to rewrite it in terms
of the topological charge plus a positive definite contribution,
\begin{eqnarray}
\frac{{\cal E}}{L} = \int {\rm d}^2x
\left\{\frac{8}{g^2}\left|\frac{\partial_{\bar\zeta}\phi}{1+\bar\phi
\phi}
\right|^2 + \frac{1}{g^2}\varepsilon_{\mu\nu}\partial_\mu
\left(\frac{\bar\phi\,  i\stackrel{\leftrightarrow}{\partial_\nu}
\phi}{1+\bar\phi
\phi}
\right)
\right\}\, ,
\label{rttc}
\end{eqnarray}
where the second term, the integral over the full derivative,
presents the topological charge
and the integral runs in the plane transverse
to the string.
Instantons saturate the topological charge; since
$\pi_2 (S_2) = Z$, the saturated solutions are labeled by an
integer
$n$, equal to the topological charge. The surface term contribution in
Eq.
(\ref{rttc}) is thus proportional to $g^{-2} n = \xi n $.

In four dimensions the instantons
present the BPS saturated strings. These strings are rather peculiar.
Since the two-dimensional theory is classically (super)conformally
invariant, the two-dimensional instantons can have any size
(correspondingly,
the cross section of the string in four-dimensional theory
can  be arbitrary). The larger is the
transverse size of the string the
smaller is the energy density in the string. However, the
string tension remains constant
proportional to $g^{-2} =\xi$.
This is the limiting profile of the Abrikosov string in SQED
with the Fayet-Iliopoulos term
-- the profile it acquires when the vector field mass tends to infinity
while the remaining degrees of freedom of the
matter fields remain massless.

For our purposes it is important to interpret
the surface term contribution in the
string tension   in terms of the
 $(1/2,1/2)$ central charge
of the four-dimensional SQED. Upon  inspecting Eq. (\ref{tfit})
we  conclude that this contribution comes from the
first term in Eq.  (\ref{tfit}). The field $A_\mu$ is not dynamical in
the
limit under consideration, and is expressible in terms of
the residual scalars. Since our consideration
is quasiclassical, it is not surprising that the current of the  matter
fermions does not contribute. The second term in Eq. (\ref{tfit})
does not contribute either -- as was discussed,
its coefficient is ambiguous.

The O(3) (or $CP_1$) model belongs to a
more general class of  $CP_N$   models.
The  latter can be derived as the low-energy limit of SQED
with the FI term and  with $N+1$ chiral
matter superfields (all of them have charge $+1$),
in the limit $e^2\to\infty$.  One can eliminate the nondynamical
$A_\mu$ field, much in the same way as in $CP_1$,
arriving in this way at a nonlinear sigma model.

One has to introduce complex coordinates
$w_{i}^j={\phi_i}/{\phi_j}$ where $i\neq j$  which
can be considered as the scalar components of the chiral
superfields $\Phi_{i}^{j}$. The action can be written
in terms of $\Phi_{i}^{j}$  as follows
\beq
S = \frac{1}{2g^2} \int {\rm d}^4 x  {\rm d}^2\theta
 {\rm d}^2\bar\theta\, \ln \left (1+\sum_{i,j}\bar\Phi_{i}^{j}
\Phi_{i}^{j}
\right)\,.
\eeq
The identification $\xi={1}/{g^2}$ is transparent
since both parameters determine the size of
the target manifold in two formulations.

The general expression for the central charge is  \cite{3dwj}
\beq
Z= \int {\rm d}^2x
\left\{ \partial_\zeta \left(
K_{\phi}\partial_{\bar{\zeta}}\phi-
K_{\bar{\phi}}\partial_{\bar{\zeta}}\bar{\phi}\right)+
\partial_{\bar{\zeta}}(K_{\bar{\phi}}\partial_{\zeta}
\bar{\phi}-K_{\phi}\partial_{\zeta}\phi)
\right\}\,,
\eeq
where the complex variable $\zeta$ is defined in Eq. (\ref{zetabar})
the subscripts $\phi,\,\,\bar\phi$ denote the $\phi,\,\,\bar\phi$
partial derivatives of the K\"ahler metric.

More generally, we expect similar strings
for all toric varieties which can be presented
as low energy limits of gauged linear sigma model.
In Sec. 9 we shall encounter one more  example of the
K\"ahler  sigma model coupled to the
Abelian gauge field -- the low-energy effective action
for ${\cal N}=2$ SUSY Yang-Mills theory in four dimensions.

\section{Supersymmetric gluodynamics}

To begin with, consider the simplest non-Abelian gauge model, SUSY
gluodynamics. The Lagrangian is
\begin{equation}
{\cal L} =  \frac{1}{4g^2}  \int\!{\rm d}^2\theta \,\mbox{Tr}\, W^2 +
\,
\mbox{H.c.}
\, ,
\label{SFYML}
\end{equation}
where
$ W = W^aT^a$, and $T^a$ are the generators of the gauge group $G$
in
the
fundamental representation.  Although the
gauge group $G$  can be arbitrary, for definiteness
we limit ourselves to SU($N$). In components
\begin{equation}
{\cal L} =  \frac{1}{g^2} \left\{ -\frac 14
G_{\mu\nu}^aG^{a\mu\nu} +
i\lambda^{a \alpha}
{\cal D}_{\alpha\dot\beta}\bar\lambda^{a\dot\beta}
\right\} \,  .
\label{SUSYML}
\end{equation}
There is a supermultiplet of  the classically conserved currents
(for a recent review see e.g. \cite{SVr}),
\begin{eqnarray}
{\cal J}_{\alpha\dot\alpha} &\!\!=&\!\! -\frac{4}{g^2}\,\mbox{Tr}
\left[e^V
W_\alpha e^{-V}\bar W_{\dot \alpha}\right]\nonumber\\[0.2cm]
 &\!\!=&\!\!  R_{\alpha\dot\alpha} -
\left\{  i\theta^{\beta} J_{\beta\alpha\dot\alpha}
 +
\mbox{H.c.}    \right\} -
 2\, \theta^{\beta}\bar{\theta}^{\dot\beta}  \,
     J_{\alpha\dot\alpha\beta\dot\beta} +\dots\;,
\label{decom2}
\end{eqnarray}
where $R_{\alpha\dot\alpha} $
is the chiral current,
$J_{\beta\alpha\dot\alpha} $ is the
supercurrent,  and ${J_{\alpha\dot\alpha\beta\dot\beta}}$ is a
combination of the   energy-momentum tensor
$\vartheta_{\alpha\dot\alpha\beta\dot\beta}=
(\sigma^\mu)_{\alpha\dot\alpha}(\sigma^\nu)_{\beta\dot\beta}\,
\vartheta_{\mu\nu}$
and a full derivative appearing in the central charge, namely,
\begin{eqnarray}
R_{\alpha\dot\alpha} &\!\!=&\!\!
-
\frac{4}{g^2}\,\mbox{Tr}\lambda_\alpha\bar\lambda_{\dot\alpha}\,
,
\nonumber\\[0.2cm]
J_{\beta\alpha\dot\alpha}&\!\!=&\!\!
(\sigma^\mu)_{\alpha\dot\alpha}
\,
J_{\mu\,,\beta}=\frac {4i}{g^2}\,\mbox{Tr}\, G_{\alpha\beta}\,\bar
\lambda_{\dot \alpha}\,,\nonumber\\[0.2cm]
J_{\alpha\dot\alpha\beta\dot\beta}&\!\!=&\!\!
\vartheta_{\alpha\dot\alpha\beta\dot\beta}
-\frac{i}{4}\varepsilon_{\alpha\beta}\partial_{\gamma\{\dot\beta}
\,
R^\gamma_{\dot\alpha\}}+
\frac{i}{4}\varepsilon_{\dot\alpha\dot\beta}\partial_{\dot\gamma
\{\beta}
\,
R^{\dot\gamma}_{\alpha\}}
\,,\nonumber\\[0.2cm]
\vartheta_{\alpha\dot\alpha\beta\dot\beta}
&\!\!=&\!\!
\frac{2}{g^2}\,\mbox{Tr}\left[i\,\lambda_{\{\alpha
}{\cal
D}_{\beta\}\dot\beta} \bar \lambda_{\dot \alpha}
-i\,\left(
{\cal D}_{\beta\{\dot\beta}\lambda_\alpha \right) \,\bar
\lambda_{\dot
\alpha\}}
+ G_{\alpha\beta}
\bar G_{\dot\alpha\dot\beta}\right]\,.
\label{arcr}
\end{eqnarray}
The symmetrization over $\alpha,\beta$ or $\dot \alpha,\dot\beta$
is marked by the braces. In fact, since the chiral current is
classically conserved
(so far we disregard anomalies), symmetrization in the third line is
superfluous: the corresponding expressions are automatically
symmetric.
To obtain the expression on the right-hand side from
$\mbox{Tr}
\left[e^V
W_\alpha e^{-V}\bar W_{\dot \alpha}\right]$
we observe
that the expression for $J_{\alpha\dot\alpha\beta\dot\beta}$
has mixed symmetry: the part symmetric in $\{\alpha \, ,\beta\}$
{\em and}  $\{\dot\alpha \, ,\dot\beta\}$ is the $(1,1)$ Lorentz
tensor, it represents
 the (traceless) energy-momentum
tensor. The remainder, i.e.
the part  symmetric in $\{\alpha \, ,\beta\}$
and antisymmetric in $\{\dot\alpha \, ,\dot\beta\}$ or
 {\em vice versa}, is the
$(0,1)+ (1,0)$ Lorentz tensor.
The part antisymmetric in both $\{\alpha \, ,\beta\}$
{\em and}  $\{\dot\alpha \, ,\dot\beta\}$ is $(0,0)$.
It represents the traces which vanish in the classical approximation.
It is quite obvious that the only part relevant for the central charge
is $(0,1)+ (1,0)$ piece in $J_{\alpha\dot\alpha\beta\dot\beta}$.
This means, in particular, that the inclusion of the traces
will have no impact on the central charge.

It is easy to see that
\beq
\left\{ Q_\gamma\, , \bar J_{\dot\beta\dot\alpha\alpha}\right\}
=2 J_{\alpha\dot\alpha\gamma\dot\beta}\, .
\label{safd}
\eeq
Combining this equation with the third line in Eq. (\ref{arcr}) we
conclude that the centrally
extended algebra is given by Eq. (\ref{bsa})
with
\beq
a^\nu = \frac{1}{4} \, R^\nu\, .
\label{okon}
\eeq

Unlike the central extension relevant for the domain walls, which
appears \cite{two} -- \cite{four} as a quantum anomaly,
in the problem at hand the algebra gets a full-derivative term at the
tree level. The presence of the anomaly manifests itself
through the fact that the energy-momentum tensor ceases to be
traceless, and $\partial_\nu R^\nu$ no more vanishes. On general
grounds it is clear, however, that Eqs. (\ref{bsa}),  (\ref{okon})
stay intact.

The occurrence of a full-derivative term in the algebra
presents a precondition for a nontrivial central extension.
Whether or not this term actually vanishes is a dynamical issue
which depends on the presence of the string-like solitons.
These may be strings, or domain-wall junctions, as in
Ref. \cite{3dwj}.  SUSY gluodynamics
is a strongly coupled theory;
therefore, one cannot use
quasiclassical considerations to search/analyze
solitons. The hope is that there is a dual description
in terms of effective degrees of freedom, for which quasiclassical
analysis may be relevant. Within this dual description
the second term in Eq. (\ref{bsa}) is mapped onto some relevant
operator of the effective theory.
 It is clear that the second term in Eq. (\ref{bsa})
is the  necessary but not sufficient condition for the
existence of the saturated strings.
If it were absent, there would be no
hope.

\section{Generic Non-Abelian Model with Matter}

The $(1/2,1/2)$ central charge
in the generic non-Abelian theory is obtained by combining
the expressions we have derived in the previous sections.
The operator $a_\mu$ in Eq. (\ref{bsa}) receives contributions from
the gluino
term, as in Sec. 6, which is unambiguous,
and the contributions from  matter
(both, the scalar and spinor components of matter enter),
as in the generalized Wess-Zumino model (Sec. 3), whose coefficients
are not fixed -- they depend on how one defines
the supercurrent in those terms that are total derivatives.
This ambiguity derives its origin from that in the definition of the
supercurrents,
\begin{eqnarray}
J_{\alpha\beta\dot\beta} &\!\!=&\!\!  \frac{2}{g^2}
\left(iG^a_{\beta\alpha} \bar\lambda^a_{\dot\beta}
+ \epsilon_{\beta\alpha} D^a\bar\lambda^a_{\dot\beta}
\right) + 2\sqrt{2}\sum\left[
\left({\cal D}_{\alpha\dot\beta}\phi^\dagger
\right)\psi_\beta -i\epsilon_{\beta\alpha}
F\bar\psi_{\dot\beta}\right]
\nonumber\\[0.2cm]
&\!\!- &\!\! \frac{\sqrt{2}}{3}\sum \left[
\partial_{\alpha\dot\beta}(\psi_\beta \phi^\dagger)
+\partial_{\beta\dot\beta}(\psi_\alpha \phi^\dagger)
-3\epsilon_{\beta\alpha}\partial^\gamma_{\dot\beta}(\psi_\gamma
\phi^\dagger)
\right]
\, ,
\label{scsqcd}
\end{eqnarray}
where the sum runs over all matter supermultiplets, $D^a$ and $F$
are the corresponding $D$ and $F$ terms.
The second line is conserved by itself, nondynamically;
the spatial integral of the time-like component
reduces to the integral over the total derivative for the second line.
Therefore, it may or may not be included
in the definition of the supercurrents.
This is the supersymmetric analog of the ambiguity
in the energy-momentum tensor in nonsupersymmetric theories
with the scalar fields. The ambiguity in the choice
of $J_{\alpha\beta\dot\beta} $ leads, with necessity,
to the fact that the coefficients of the matter terms in $a^\mu$ in Eq.
 (\ref{bsa}), namely, $a^\mu_{(\psi )}$ and $a^\mu_{(\phi )}$,
are not uniquely fixed.

Due to this ambiguity, the matter component of
$a^\mu$ cannot contribute to $Z$ for strings (it could contribute,
though,
for the wall junctions and other similar object with the axial
geometry).

\section{Strings
Cannot be Saturated in ${\cal N}=1$ Non-Abelian Gauge Theories in
Weak  Coupling}

Here we will prove that
in the absence of the U(1) factors, even if the theory
under consideration does support string-like solitons
in the quasiclassical consideration (some examples are discussed
e.g. in Ref. \cite{Strass}), the central charge vanishes with necessity.
Therefore, these strings {\em cannot be saturated}.

In weak coupling (i.e. for the string solitons
in the quasiclassical treatment)
the $(1/2,1/2)$ central charge  must be saturated
 by the term with the {\em bosonic}
axial current. (We remind that the FI term is absent).
As was explained, the coefficient of this term is
not unambiguous -- it depends on the definition of the
supercurrent (e.g. minimal {\em versus} conformal).
Since we are interested in the
string solitons, rather than the wall junctions,
this ambiguity cannot be canceled by that in the $(1,0)$
central charge, since the latter must identically vanish.
This is dictated by the Lorentz symmetry arguments.
This means that the $(1/2,1/2)$
central charge must vanish identically.

 The consideration above
shows that if  the BPS  objects with the axial geometry
exist in the quasiclassical limit (in non-Abelian
gauge theories), the stringy core must be
accompanied by objects with the $(1,0)$ charges.
In four dimensions
domain walls do the job.   Within the brane picture
it is possible to consider four-dimensional
theories as that  on the brane embedded in $M$ theory.
For instance, the expected domain wall junctions
in ${\cal N}=1$ Yang-Mills theory -- the
gauge analog of the junctions in the GWZ models --
can be identified as a
junction of  M5 branes, so that  the definition
of the  current for the theory on M5
removes any ambiguity.

\section{Strings in the Seiberg-Witten ${\cal N }=2$ Model }

Here we will speculate on possible BPS strings at strong coupling.
As we already know, such strings do not appear in weak coupling.
The $(1/2,1/2)$ central charge
(appearing in the anticommutator $\{ Q,\, \bar Q\}$)
is not holomorphic -- it need not depend holomorphically
on the chiral parameters, in contradistinction with the $(1,0)$ charge.
This means, that even if both the weak and strong coupling regimes
are
attainable in one and the same theory,
generally speaking,  nothing
can be said regarding the BPS strings in  the strong coupling regime
from the behavior at weak coupling.

\subsection{Strings in Pure ${\cal N }=2$ Yang-Mills Theory}

Turn now to discussion of the  ${\cal N }=2$ Yang-Mills theory
without matter hypermultiplets. The
exact solution for the low-energy effective
action, as well as the exact  spectrum of the BPS  particles,
are known \cite{ws}. Now
we address the  issue of  possible stringy central
charges, besides the standard  ones,
saturated by particles \cite{wo}.
>From the discussion above we saw that it can be
attributed only  to the gluino  axial current since
there is no  FI term in the model. Let us
 restrict ourselves to SU(2) gauge
group.

The key features of the Seiberg-Witten
solution can be summarized as
follows.
The vacuum manifold develops the Coulomb branch which is
parametrized by  the global coordinate, the order parameter
$u=\langle {\rm tr}\phi^2\rangle$.  At low energies
the effective theory becomes Abelian and
is described by a single holomorphic function -- prepotential
$\cal {F}$ which determines the effective coupling constant
of the theory $\tau={\partial^2 \cal{F}}/{\partial a^2}$,
as well as the K\"ahler metric on the Coulomb
branch of the moduli space, which appears to be a
one-dimensional special K\"ahler
manifold. The K\"ahler potential
can be found from the prepotential as follows
\beq
K(a,\bar  a)={\rm Im} a_{D}\bar a
\eeq
where $a$ is the vacuum  value of the third component of the
scalar field and
$a_{D}={\partial \cal{F}}/{\partial a}$. The variable $a$ can be
expressed in terms of variable $u$ as follows:
\beq
a(u)=\int_{-\sqrt{u+\Lambda^2}}^{-\sqrt{u-\Lambda^2}}
\frac{x^2 dx}{\pi\sqrt{(x^2-u)^2-\Lambda^4}}\, .
\eeq

Unlike the variable $u$, the variable $a$ cannot be considered as a
global
coordinate on the moduli space since
the K\"ahler metric ${\rm Im}\tau(a)$ has zeros
(here $\tau$ is the complexified
coupling constant). Therefore,
to analyze the  complex plane of $a$, an explicit expression for $a(u)$
is
needed. A direct inspection shows that the region of small $a$
is essentially removed from the
complex plane so that  $|a(u)|> {\rm const}\,  \Lambda$.

The lower bound  on $a$  can be seen also geometrically
if we recall that it is just the mass of the $W$ boson, which can be
represented in the theory on D3 probe
as the pronged string connecting the
probe and the split O7 orientifold \cite{probe}.
It is clear that the
minimal mass of the $W$ boson geometrically is the distance
between
the
7-branes on the $u$ plane; it is, thus,
proportional to $\Lambda$. Therefore,
we see that $\pi_1$ of the scalar field  manifold is nontrivial --
topologically stable objects with the axial geometry are expected,
provided  $a$
winds around the ``forbidden" region.

Whether these objects are strings (i.e. have finite energy per unit
length)
depends on dynamics, on how fast the volume energy density
dies off as we go away from the axis in the perpendicular
direction. The convergence could be ensured by the
appropriate form of the K\"ahler metric, as in the sigma models.
It is quite obvious, that in this case
 the string tension
\beq
T={\rm const}\, \Lambda^2\, .
\eeq

The existence of
such the stable objects with the axial geometry  would be a purely
strong coupling effect since at the classical level the
point $ a=0$  is attainable, and, correspondingly,
$\pi_1$ is trivial.  If the strings do exist, they may be BPS-saturated
provided the term due to the gluino current $R^\mu$ in the central
charge is
nonvanisihing.
To this end the gluino current must fall off at large distances
$r$ from the axis as $1/r$. Finiteness of the string tension
would imply then that effective degrees of freedom coupled to
$R^\mu$
form a U(1) gauge interaction. If the string tension is finite and
the gluino current falls off at large distances
$r$ from the axis faster than  $1/r$, the string is tensionless.

\subsection{Strings in ${\cal N}$=2 SQCD}

Adding the matter hypermultiplets to the model discussed in Sec. 9.1
we get  ${\cal N}$=2 SQCD.
 Since
there is no restoration of the SU(2) gauge symmetry
at the  generic point at the Coulomb branch,  the
``forbidden" region on the complex $a$ plane exists in the
theory with the fundamental matter too. The
 BPS strings may appear on the Coulomb
branch,  with the  tension  saturated
by the  $R$ current of gluinos. The
tension now depends on the masses of the
fundamental matter  and can be determined, in principle,
from the explicit expression for $a(u, \Lambda, m)$.

Moreover,  the  Higgs branch (parametrized by
the vacuum expectation values of the fundamentals
$\langle Q\rangle , \,\, \langle\tilde Q\rangle $) is possible, and
the question of the  BPS strings on the Higgs
branch can be addressed. We recall that geometrically
the  Higgs branch is the  hyper-K\"ahler manifold \cite{aps}
(for a review see  \cite{ap}) whose metric can be determined
classically.
It  is not renormalized by quantum corrections. Actually,
the Higgs branch for SU$(N_c)$ theory with $N_f$
flavors  is the cotangent bundle of the Grassmannian
$T^{*}{\rm Gr}_{N_c,N_f}$,  with the antisymmetric $N_c$\,-form.
The metric on this manifold can be found from the K\"ahler potential
\beq
K(Q,\tilde Q)={\rm Tr}\sqrt{k^2+MM^{\dagger}}\, ,
\eeq
where $k$ is a solution of the equation
\beq
{\rm det}\, \left( k 1_{N_f} +\sqrt {k^2 1_{N_f}+MM^{\dagger}}
\right) = {\rm det}\,(QQ^{\dagger})\,,
\eeq
and $M=Q\tilde Q$ is the meson matrix.

Since $\pi _{2}({\rm Gr}_{n,k})\neq 0$,  instantons
in the two-dimensional sigma model on $T^{*}{\rm Gr}_{n,k}$
are possible. The  arguments presented in  Sec. 5 suggest that these
instantons can be interpreted as strings on the Higgs branch.
It would be interesting to understand whether a version of the
string on the Higgs branch recently found in \cite{yung}
can be BPS saturated.

The existence of the BPS string on the Higgs branch was
recently conjectured within the brane approach \cite{oz}.
This string was expected to be tensionless at the
root of the Higgs branch, which qualitatively
agrees with the discussion above.

\subsection{Softly broken ${\cal N}$=2 theory
(strings in ${\cal N}=2$ SQED)}

If the softly broken ${\cal N}=2$ Yang-Mills theory
is considered near the monopole or dyon singularities
the effective low-energy theory which ensues is
${\cal N}=2$ dual SQED. This is the famous Seiberg-Witten
result. A small mass term of the chiral superfields
of the original ${\cal N}$=2  non-Abelian theory
is translated
in a small perturbation of the superpotential for the
matter fields in SQED.
If the monopole (or dyon) superfields
are denoted as $M,\, \tilde M$
 the
superpotential in the low-energy  SQED
can be written as
\beq
{\cal W}=\mu\,  u(a_{D}) +\tilde{M}a_{D}M\, ,
\label{ddd}
\eeq
where $a_D$ is a chiral superfield which is
the ${\cal N}=2$ superpartner of the (dual) vector superfield.
The second term in Eq. (\ref{ddd})
is fixed by ${\cal N}=2$ supersymmetry.
The parameter
$\mu$ in the first term is small. Generically
$\mu\,  u(a_{D}) $ breaks ${\cal N}=2$ supersymmetry
down to ${\cal N}=1$. However, in the linear approximation,
when
\beq
{\cal W}=\mu a_{D} \Lambda +\tilde{M}a_{D}M\, ,
\label{fff}
\eeq
${\cal N}=2$ is unbroken.

Let us forget about the origin
of ${\cal N}=2$ SQED and discuss this U(1) theory
with the superpotential (\ref{fff}) {\em per se}.
Minimization of the potential stemming from
(\ref{fff}) yields
the monopole condensation.
The Abrikosov strings obviously do exist.
Their tension is
proportional
to  $\mu$.
They were discussed in the literature previously~\cite{DS,HSZ}.
The classical equations for the string reduce
to Eq. (\ref{sateq}). \footnote{It should be taken
into account that on the solution
$ | M| = |\tilde M|$.}
Thus, the string is
saturated. The question is how this could happen
given that the $(1/2, 1/2)$ central  charge must vanish
in the absence of the FI term.

The central charge in the anticommutator
$\{Q_\alpha \bar Q_{\dot\alpha} \}$
is indeed zero. One should not forget however,
that SQED with the superpotential (\ref{fff})
is an ${\cal N}=2$ theory --
there exist two supercharges $Q, \, Q '$
of the type $(1/2,0)$
and two supercharges $\bar Q, \,\bar Q '$
of the type $(0,1/2)$. Therefore, one should look
for the central extension in  the anticommutator
of the general form
$\{ {\cal Q}_\alpha \bar{\cal Q}_{\dot\alpha} \}$
where ${\cal Q}$ is a linear combination
of $Q$ and $Q '$.  A nonvanishing cenral term
of this type does exist.

If we now return to the
original  non-Abelian ${\cal N}$=2 theory,
we  conclude that at
small $\mu$ the string is (approximately) BPS saturated.
It becomes exactly saturated in the limit
$\mu\to 0\,,\quad \Lambda\to\infty$
with $\mu\Lambda$ fixed~\cite{DS,HSZ}.
The saturation is approximate, rather than exact, since
higher order terms in $\mu$ (non-linear in $a_D$ terms in the
superpotential
of the low-energy U(1) theory)
break ${\cal N}=2$
 and
return us back to the ${\cal N}=1$ theory.
In ${\cal N}=1$ the extra supercharges $Q'
, \, \bar Q '$ disappear, while the
central charge in the anticommutator
$\{Q_\alpha \bar Q_{\dot\alpha} \}$
vanishes.

\section{The Brane Picture: How It Corresponds to Field Theory}

\subsection{The Fayet-Iliopoulos string as a membrane}

With the brane picture in mind, we
can look for the brane configuration corresponding to
the BPS strings  discussed above. The interpretation of the
strings
whose tension is proportional to the four-dimensional
FI terms is rather simple.
Let us consider the brane configuration relevant for the Abelian
${\cal N}=2$
Yang-Mills
theory in the IIA picture. It consists of the pair of the parallel
NS5 branes with the worldvolumes $(x^0, x^1, x^2, x^3, x^4, x^5)$,
plus
a single D4 brane with the worldvolume $(x^0, x^1, x^2, x^3, x^6)$.
The gauge theory is defined on the worldvolume of the  D4 brane,
and the
distance between the NS5 branes along the $x^6$ direction
plays the role of the
inverse coupling in the Abelian theory.
Since the four-dimensional FI terms have the meaning of the relative
distance between the
NS branes in  $(x^7, x^8, x^9)$  \cite{GK},
the
``FI strings" are nothing but the D2 branes stretched between the NS
branes in some of $(x^7, x^8, x^9)$ directions. The rest of their
worldvolume
coordinates coincide with the D4 ones.

This picture gets slightly modified if one considers
the Abelian  ${\cal N}=1$  theory. According to
the well-known procedure (see, for instance, \cite{GK}), one has then
to rotate one of the NS5 branes, which now has
$(x^0, x^1, x^2, x^3, x^8, x^9)$ as the
worldvolume.  The Fayet-Iliopoulos term
has now the meaning of the displacement of the  NS5 branes along
$x^7$.
The D2 brane stretched between the  NS5 branes with the
worldvolume
$(x^0, x^1, x^7)$
plays the role of the BPS string.

Let us note  that the FI string can be elevated  smoothly in the $M$
theory.
Indeed,  the NS5 branes and the  D4 brane can be identified with the
single
M5 brane in the $M$ theory. The FI string
can be considered as an M2 brane
stretched between two components of the M5 branes.
The tension of the
FI string is proportional to the length of the  M2
brane along  the $x^7$ direction
and, therefore, proportional to the value of the FI parameter $\xi$, in
full
agreement with the field theory expectations. Recently a
similar picture for the FI strings was discussed in \cite{oz}.

\subsection{On  (conjectured) strong coupling BPS strings
via
\\  branes}

The BPS saturated objects with the axial geometry were
discussed in the brane picture previously. For instance,
the  domain wall junction in ${\cal N}=1$ supersymmetric
gluodynamics
which is expected to saturate both, the $(1/2, 1/2)$ and $(1,0)$
central  charges, occurs as the intersection
of the M5 branes,  since
the domain walls were identified as the M5
branes wrapped on 4-manifold in $M$ theory \cite{wittenmqcd}.

Here we would like to add a few remarks on a
possible interpretation of the strong coupling
BPS strings in the brane picture.
Previous attempts to recognize BPS tensionless
string in four dimensions, apparently  seen within the brane
approach \cite{HK}, were based on the intersection
of the M5 branes
or the  M2 brane stretched between two M5 branes.
The BPS string on the Coulomb branch discussed in Sec.~9
is nothing but a wrapped M5 brane,  since its
tension is proportional to the area on the region on the Coulomb
branch.
However,  the explicit geometry of
intersection of the M5 branes yielding saturation
of the $(1/2,1/2)$  central charge   is still to be clarified.

Another possible approach to the brane interpretation of
the  BPS strings   in the Yang-Mills theories
follows from the correspondence between  the Yang-Mills theories
in four dimensions  and two-dimensional sigma models.
It was recently recognized
\cite{hh,d,dht} that there is a close relation
between the two-dimensional  $CP_N$ models (which have ${\cal
N}=2$)
and the Yang-Mills theories in four dimensions, with ${\cal N}=1$
or ${\cal N}=2$,
with or without fundamental matter. In the latter case
the correspondence  relies on the coincidence
of the  spectra of the BPS domain walls in four
dimensions  and BPS solitons in two.

A more direct relation connects the ${\cal N}=2$ theory with $N_f$
flavors
at the root of the baryonic branch of the
moduli space with the   $CP_{2N_c-N_f-1}$ model
\cite{d,dht}.  The translation dictionary between the two
models looks as follows: the complex coupling
in four dimensions  corresponds to a complex parameter
combining the two-dimensional
 FI term with the $\theta$ term; twisted mass terms
in $d=2$ correspond to the coordinates on
the Coulomb branch in the $d=4$ theory; finally,
the Riemann surfaces providing the BPS
spectra in both theories are the same.

The correspondence above has a rather simple
explanation in  the brane description of both theories. It appears that
the brane configurations for both theories are actually the same.
The  $d=4$ theory is defined on the worldvolume
of the D4 branes stretched between a  pair
of the NS5 branes. The coupling constant is just the distance between
the NS5 branes. In $M$ theory all branes above are elevated
 to  a pair of the  M5 branes,
one of which is flat and the second is wrapped around the Riemann
surface. The configuration is described by the holomorphic
embedding into four-dimensional space
\beq
(t-\Lambda^N)
\left\{ t\Lambda^{N-N_f}\prod ^{N-N_f} (v-\tilde{m_i}) -
\prod^N (v-m_i)\right\}
=0\,,
\eeq
where the first factor represents the flat brane, while the second
the curved one,  and  $m_i$'s correspond to the masses
of the fundamental hypermultiplets.

Let us add a D2 brane and consider the Abelian
gauge theory on its worldvolume. If the D2 brane is stretched
between the same
NS5 branes we arrive at the $CP_N$ model in $d=2$ where it has
the extended supersymmetry,
${\cal N}=2$. This explains
the coincidence between the complexified coupling constant in
the four-dimensional  theory and the
FI term in the two-dimensional  theory. Therefore,  the picture can
be
apparently
interpreted as follows:  the $d=2$  sigma model,
 with the twisted masses added,
is the theory on the brane which is the
probe for the ${\cal N}=2$   low-energy theory  in four dimensions.

In \cite{d,dht} it was shown  that the spectrum
of the BPS particles in ${\cal N}=2$ theory
at the root of the baryonic branch exactly coincides with
the spectrum of BPS dyonic
kinks in the corresponding $CP_N$ model.
Moreover, the brane identification shows that the hypermultiplets
in $d=4$ and $d=2$ arise essentially  in the same way. Therefore, we
can
use the relation between the models in a two-fold way. The existence
of instantons in the $CP_N$ model implies that one can expect
BPS saturated strings at the root of the baryonic branch.
In the opposite direction, the  $(1/2,1/2)$ central  charge in
four dimensions  can be mapped  into the central
charge of the ${\cal N}=2$ two-dimensional theory.
Since in the  formulation of the
sigma model with the nondynamical vector  field,
 the gauge 2-potential plays the role of the current
$a^\mu$ in Eq. (\ref{bsa}), the  central charge is actually mapped
onto the Chern number $\int Adx$. Certainly, these issues
need further clarification. We hope to discuss them
elsewhere.

\section{Conclusions and Discussion}

In this paper we elaborated the generic structure of
of the central charges in  supersymmetric gauge theories in
four dimensions. The central finding is that the $(1/2,1/2)$
charge is ambiguous in the part related to the matter fields,
due to possible total derivative terms in the supercurrents.
The part related to the gauge fields (including gaugino)
is unambiguous. That is why in the weak coupling regime
the only model admitting BPS strings is SQED with the
Fayet-Iliopoulos term. In the non-Abelian theories
the
Fayet-Iliopoulos term is forbidden; hence, at weak coupling there can
be no
BPS strings. Even if some strings exist, they are nonsaturated
with necessity.
 These assertions are  proven at the theorem level.

The ambiguity we found does not preclude from existence
other BPS saturated objects with the axial geometry, i.e. the wall
junctions.
The ambiguity in the $(1/2,1/2)$ charge is combined with that
in the $(1,0)$ charge to produce a well-defined answer
for the tension of the walls and the ``hub" in the middle.
We presented some examples.

The strong coupling regime is a different story.
Since the analyticity argument does not apply
to the $(1/2,1/2)$ charge, the existence/non-existence of the
BPS strings should be discussed separately
at weak and strong couplings -- the lessons we learn at weak
coupling say nothing about possible scenarios at strong coupling.
We speculated on  different cases when the
BPS-saturated objects with axial geometry may  appear in the  strong
coupling regime. We argued that saturation
of   the $(1/2,1/2)$ charge at  strong coupling  can be attributed
to the M5 brane intersection.
(The  Fayet-Iliopoulos  BPS strings come from the M2 branes.)
The BPS-saturated strings may be expected in
the ${\cal N}=2$ Yang-Mills theories
on the Coulomb branch.

In the ${\cal N}=1$ gauge  theories
an obvious candidate for the  BPS saturation  is the domain wall
 junction. One cannot assert at the moment with absolute certainty
that the  strong coupling junction
 exists  in ${\cal N}=1$ supersymmetric
gluodynamics,  but the $M$ theory arguments
 suggest that such junctions do exist.
Additional support in favor of this conclusion is
provided by field-theoretic models considered in
\cite{GabShi}.

A comment is in order regarding the situation in supergravity
coupled to  the Yang-Mills theory.  Upon inspecting the (1/2, 1/2)
central
charge one finds the term $H=dB-K$
in the anticommutator $\{Q,\,\bar Q\}$, where $B$ is the
two-form   field and $K=AdA-\frac{2}{3}A^3$
is the  dual of the Chern-Simons current
in the Yang-Mills theory. Therefore, we see that the  axial
current of gluons enters into
the central charge,  if gravity degrees of freedom
are taken into account. We plan to  discuss this point in more detail
  elsewhere.

Since the ${\cal N}=2$ Yang-Mills
 theory enjoys duality, one can pose a question of
the duality partner of the BPS string.
Four-dimensional  BPS strings can be viewed as objects
dual (in the Dirac sense) to   localized objects.
 Indeed, since in
$d$ space-time dimensions the $ p$ brane is dual to a $(d-p-4)$
brane,
the Dirac quantization condition amounts
to the observation that the $(d-p-4)$ brane
is weakly coupled if  the $p$ brane
is strongly coupled and {\em vice versa}. Therefore, one can expect
that within the framework of duality
the strongly coupled BPS string
has something to do with the instantons at  weak coupling.

To make a conjecture regarding the central charge dualizing the
stringy one,
 let us observe  that there is a contribution in the
central charge for $\{Q,\bar Q\}$ in six dimensions,
saturated by instantonic strings. In five dimensions the
instanton presents a particle with mass ${1}/{g^2}$,
saturating \cite{seiberg} the central
charge $\int d^4 x \tilde{F}F$. In four dimensions we can expect a
remnant of this central charge resulting  from
dimensional reduction.

\vspace{1cm}

{\bf Acknowledgments}

 \vspace{0.2cm}

\noindent
We would like to thank M. Strassler,
A. Vainshtein and  A. Yung  for
useful discussions.
A part of this work was done while one of the authors (M.S.) was
visiting
the
Aspen Center for Physics, within the framework of the program
{\em Phenomenology of Superparticles and Superbranes}.
A.G. thanks TPI at the Minnesota University for the  kind
hospitality.
This work was supported in part by DOE under the grant number
DE-FG02-94ER408. The work of A.G. was also supported  by  the
INTAS
grant number
INTAS-97-0103.


\begin{thebibliography}{99}

\bibitem{1}
For a review and list of references see
M. Shifman, in  {\em Particles, Strings and Cosmology},  Proceedings
of
PASCOS 98,
   Ed.  P. Nath  (World Scientific, 1999) [hep-th/9807166].

\bibitem{two}
G.~Dvali and M.~Shifman,
{\it Phys. Lett.} {\bf B396}, 64 (1997); Erratum {\bf B407} 452,1997
[hep-th/9612128].

\bibitem{three}
A.~Kovner, M.~Shifman and A.~Smilga,
{\it Phys. Rev.} {\bf D56}, 7978 (1997)
[hep-th/9706089].

\bibitem{four}
B.~Chibisov and M.~Shifman,
{\it Phys. Rev.} {\bf D56}, 7990 (1997); Erratum {\bf D58}, 109901,
(1998)
[hep-th/9706141].

\bibitem{deAzcarraga:1989gm}
J.A.~de Azc\'{a}rraga, J.P.~Gauntlett, J.M.~Izquierdo, and P.K.~Townsend,
{\it Phys. Rev. Lett.} {\bf 63},  2443 (1989).

\bibitem{FP}
S. Ferrara and M. Porrati,
{\it Phys. Lett.} {\bf B423} (1998) 255.

\bibitem{2}
J. Edelstein, C. N\'{u}\~{n}ez, and F. Schaposhnik,
{\it Phys. Lett.} {\bf B329} (1994) 39;\\
 A.A. Penin, {\it Nucl. Phys.} {\bf B532} (1998) 83.

\bibitem{2prim}
A.A. Penin, V.A. Rubakov, P.G. Tinyakov, and S.V. Troitsky, {\it Phys.
Lett.} {\bf B389} (1996) 13.

\bibitem{3}
S.C.~Davis, A.~Davis and M.~Trodden,
{\it Phys. Lett.} {\bf B405}, 257 (1997)
[hep-ph/9702360].

\bibitem{AT}
E.R.~Abraham and P.K.~Townsend,
{\it Nucl. Phys.} {\bf B351}, 313 (1991).

\bibitem{wjone}
G.W~ Gibbons and P.K.~Townsend,
hep-th/9905196.

\bibitem{3dwj}
S.M.~Carroll, S.~Hellerman and M.~Trodden,
hep-th/9905217.

\bibitem{scsm}
S. Ferrara and B. Zumino, {\it Nucl. Phys.}
{\bf B87}, 207 (1975).

\bibitem{GabDv}
G.~Dvali, G.~Gabadadze and Z.~Kakushadze,
hep-th/9901032.

\bibitem{GabShi}
G.~Gabadadze and M.~Shifman,
hep-th/9910050.

\bibitem{dopone}
P.J.~Ruback,
Commun.\ Math.\ Phys.\ {\bf 116},  645 (1988).

\bibitem{wittenn=2}
E. Witten, {\it Nucl. Phys.} {\bf 403} (1993) 159.

\bibitem{SVr}
M.~Shifman and A.~Vainshtein,
hep-th/9902018.

\bibitem{Strass}
M.J.~Strassler,
{\it Prog. Theor. Phys. Suppl.} {\bf 131}, 439 (1998)
[hep-lat/9803009].

\bibitem{ws}
E. Witten and N. Seiberg, {\it Nucl. Phys} {\bf B426},  16 (1994);
{\it Nucl. Phys} {\bf B431},  484 (1994).

\bibitem{wo}
E. Witten and D. Olive, {\it Phys. Lett. } {\bf B78},  97 (1978).

\bibitem{probe}
O. Bergman and A. Fayyazuddin, {\it Nucl. Phys.} {\bf B531}, 108
(1998);\\
A. Mikhailov, N. Nekrasov and S. Sethi, {\it Nucl. Phys.} {\bf B531},
345 (1998).

\bibitem{aps}
P. Argyres, M.  Plesser and N. Seiberg, {\it Nucl. Phys.} {\bf B471},
159 (1996).

\bibitem{ap}
I. Antoniadis and B. Pioline, {\it Int. J. Mod. Phys.}
{\bf A12}, 4907  (1997)
[hep-th/9607058].

\bibitem{yung}
A. Yung, hep-th/9906243.

\bibitem{oz}
S. Alishahiha and Y. Oz , hep-th/9907206.

\bibitem{DS}
M. Douglas and S. Shenker, {\it Nucl. Phys} {\bf B447},
 271 (1995) [hep-th/9503163].

\bibitem{HSZ}
A. Hanany, M. Strassler, and A. Zaffaroni, {\it Nucl. Phys.}
{\bf B513},   89 (1998).

\bibitem{GK}
A. Giveon and D. Kutasov, hep-th/9802067.

\bibitem{wittenmqcd}
E. Witten, {\it Nucl. Phys.} {\bf B507},   658 (1998).

\bibitem{HK}
A. Hanany and I. Klebanov, {\it Nucl.Phys.} {\bf B482},  105 (1996).

\bibitem{hh}
A.~Hanany and K.~Hori,
{\it Nucl.  Phys.} {\bf B513}, 119 (1998)
[hep-th/9707192].

\bibitem{d}
N.~Dorey,
{\it JHEP}, {\bf 11}, 005 (1998)
[hep-th/9806056].

\bibitem{dht}
N.~Dorey, T.J.~Hollowood and D.~Tong,
{\it JHEP}, {\bf 05}, 006 (1999)
[hep-th/9902134].

\bibitem{seiberg}
N. Seiberg, {\it Phys. Lett.}  {\bf B388},   753 (1996).

\end{thebibliography}
\end{document}